\documentclass[twocolumn,pre]{revtex4}
\usepackage{amssymb}
\usepackage{latexsym}
\usepackage{graphicx}
\usepackage{dcolumn}
\usepackage{bm}
\usepackage{graphics}
\usepackage{mathrsfs}

\newcommand{\be}{\begin{equation}}
\newcommand{\ee}{\end{equation}}
\newcommand{\bea}{\begin{eqnarray}}
\newcommand{\eea}{\end{eqnarray}}
\begin{document}

\title{How a protein searches for its specific site on DNA: the role of intersegment transfer}
\date{\today}

\author{Tao Hu}
\author{B. I. Shklovskii}
\affiliation{Theoretical Physics Institute, University of Minnesota,
Minneapolis, Minnesota 55455}

\begin{abstract}
Proteins are known to locate their specific targets on DNA up to two
orders of magnitude faster than predicted by the Smoluchowski
three-dimensional diffusion rate. One of the mechanisms proposed to
resolve this discrepancy is termed ``intersegment transfer". Many
proteins have two DNA binding sites and can transfer from one DNA
segment to another without dissociation to water. We calculate the
target search rate for such proteins in a dense globular DNA, taking
into account intersegment transfer working in conjunction with DNA
motion and protein sliding along DNA. We show that intersegment
transfer plays a very important role in cases where the protein
spends most of its time adsorbed on DNA.
\end{abstract} \maketitle

\section{Introduction}

The description of how proteins interact with specific sites on DNA is of fundamental importance to molecular biology. The effectiveness of a DNA enzyme depends entirely on its ability to locate its target site quickly and reliably. It was recognized long
ago that the search by free diffusion through three-dimensional
(3D) solution is far too slow to account for the observed speed of many biological processes, and that proteins somehow arrive at their
target sites up to two orders of magnitude faster than the 3D
Smoluchowski rate \cite{Riggs, Richter}
\be J_s = 4\pi D_3bc , \label{eq:smoluchowski} \ee
where $b$ is the target radius and $D_3$ and $c$ are, respectively,
the diffusion coefficient and concentration of proteins in solution.
The idea to resolve this discrepancy goes back to Delbr$\ddot{\rm
u}$ck \cite{Delbruck}, who suggested that proteins may adsorb fairly
quickly onto a nonspecific random place on DNA and then undergo
one-dimensional (1D) sliding along the DNA strand, resulting in an
increase  of the search rate $J$ above $J_s$. Equivalently, we can
say that the average search time for the proteins $t=1/J$ falls
below the Smoluchowski time $t_s=1/J_s$. Below we characterize this
rate enhancement by the acceleration ratio $t_s/t$.

The field attracted intensive attention for many years. On the
theoretical front, the pioneering work by Berg, Winter and von
Hippel \cite{BWH} established the basis of current understanding in
this field. They showed that 1D sliding on DNA forms a kind of
``antenna" around the target site and serves to increase the
effective size of the target. This large antenna size replaces the
actual target size $b$ in Eq. (\ref{eq:smoluchowski}), resulting in
a much faster search rate. The Berg, Winter and von Hippel model
predicts that the rate at which proteins find their specific target
sites on DNA depends in a nonmonotonic fashion on the ionic strength
of the solution, which seems to be qualitatively consistent with
experiments.

\begin{figure}[htb]
\centering
\includegraphics[width=0.40\textwidth]{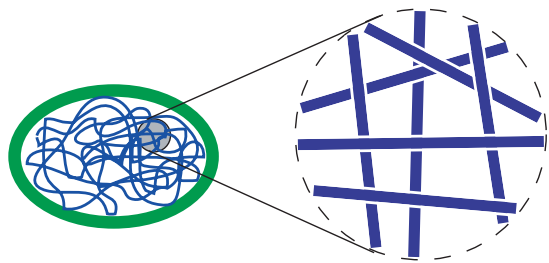}
\caption{A DNA globule. The long DNA is forced to return many times
by the wall of a prokaryotic cell. On the right, a blown-up view
shows a typical region of the transient network at a scale much
smaller than the DNA persistence length $p$. This figure represents
a very dense case, where the nearest neighbor distance between DNA
segments, each of length $p$, is shorter than
$p$.}\label{fig:globule}
\end{figure}

In recent years, the sliding mechanism has been revisited several
times \cite{Halford, Mirny}, but the question of how the protein
search time depends on DNA conformation was not addressed. It is
well known that DNA is coiled at length scales larger than its
persistence length. When the coil cannot fit in the volume
available, e.g. in the nucleoid in a prokaryotic cell, it must be a
globule, as it is forced to fold back into the volume after each
contact with the walls. Locally, the globule resembles a transient
network with a certain mesh size (see Fig. \ref{fig:globule}). A
scaling theory was recently proposed to account for the role of
different DNA conformations \cite{BPJ}. This theory deals only with
proteins with a single DNA binding site and ignores the motion of
DNA in solution. Our goal in this paper is to relax these
restrictions.

Berg, Winter and von Hippel \cite{BWH} pointed out that in addition
to 1D sliding, proteins with two nonspecific DNA binding sites may
benefit from another facilitating mechanism termed ``intersegment
transfer". Indeed, such proteins are capable of transiently binding
to two DNA segments when the segments are close in space, even if
they are well separated from each other along the DNA contour. The
subsequent segmental diffusion of DNA then disrupts these
double-bound states, resulting in the protein being transferred to a
remote position on the DNA without net dissociation of protein into
the water.

The existence of intersegment transfer in principle has been
confirmed by a number of well-designed \emph{in vitro} experiments
\cite{Fried, Lieberman, Iwahara}. These experiments measured the
dissociation rate of proteins from a prepared complex of the protein
and a short piece of specific DNA. The complex was placed in a
solution of short nonspecific DNA molecules, and the dissociation
rate was measured as a function of the concentration of nonspecific
DNA. All the proteins used in \cite{Fried, Lieberman, Iwahara},
namely lac repressor \cite{Fried}, glucocorticoid receptor
DNA-binding domain protein \cite{Lieberman} and human Hox-D9
homeodomain \cite{Iwahara} are believed to have two DNA binding
sites so that the protein-DNA complex can adsorb a second short
piece of DNA, allowing the protein to transiently form a
double-bound state with two DNA pieces. This double-bound state
breaks up quickly (faster than the dissociation of protein to water
in the prepared protein-DNA complex) and the protein has a chance to
be transferred to the newly adsorbed DNA. As a result, the
dissociation rate of the complex grows linearly with the
concentration of nonspecific DNA. This phenomenon of inter-DNA
transfer is essentially similar to the intersegment transfer of
proteins inside a single long DNA strand. Direct observation of
intersegment transfer was also achieved by a scanning force
microscopy study of the translocation of RNA polymerase in E. coli
\cite{Bustamante}.

In this paper, we propose a scaling theory of the target search time
for proteins with two DNA binding sites, which combines the effects
of 3D diffusion, 1D sliding, intersegment transfer and DNA motion.
Our main interest is the search time for the biologically relevant
case of globular DNA. However, its complex geometrical properties
combined with the several mechanisms of protein motion make this
problem very complicated. Therefore, we start from a relatively
simple case, namely the search time in a solution of short, straight
double-helix DNA molecules among which only a small fraction carry
the specific targets. In this situation we are able to include the
effects of intersegment transfer and establish connections with the
\emph{in vitro} experiments on short DNA \cite{Fried, Lieberman,
Iwahara}. Our analysis of this case is detailed in Sec. II, and a
summary of the resulting scaling regimes is shown in Fig.
\ref{fig:searchrate}.

\begin{figure}[htb]
\centering
\includegraphics[width=0.45\textwidth]{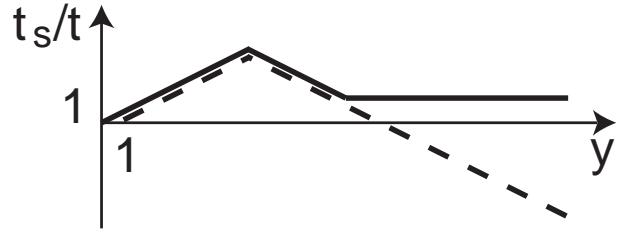}
\caption{Schematic dependencies of the acceleration rate on the
adsorption strength $y$, with (solid line) or without (dashed line)
DNA motion and intersegment transfer. Both the acceleration rate
$t_s/t$ and $y$ are given in logarithmic
scale.}\label{fig:globulerate}
\end{figure}

In Sec. III we apply the methods developed for short DNA pieces to
the case of a very dense DNA globule as shown in Fig.
\ref{fig:globule}. The acceleration rate $t_s/t$ is shown
schematically as the solid line in Fig. \ref{fig:globulerate},
plotted as a function of $y=\exp(\epsilon/k_BT)$, where $\epsilon$
is the nonspecific adsorption energy of the protein to DNA.
Experimentally, the value of $y$ can be controlled through the salt
concentration of the solution, since non-specific absorbtion of
proteins is controlled by Coulomb interaction between negative DNA
and the positive patch on the protein surface and may be screened by
salt concentration. For comparison, we also plot the result of Ref.
\cite{BPJ}, which ignores DNA motion and intersegment transfer, as a
dashed line. In the latter case, the acceleration rate grows first
with $y$ because protein binding to DNA increases the antenna size;
then the acceleration rate decays when most of the proteins are
fruitlessly adsorbed far from the target (or, in other words, every
protein spends most of the time adsorbed far away from the target).
Finally, the acceleration rate saturates and comes to a very low
plateau (not shown) when the antenna becomes as long as the DNA
itself. Hence, when DNA motion and intersegment transfer are not
accounted for, there is a very strong deceleration at large ionic
strength compared to the Smoluchowski rate. With the help of
intersegment transfer, however, the acceleration rate saturates at a
much higher level (larger than unity) because adsorbed proteins
become much more effective in target search.

In Sec. IV, we conclude with a discussion of the applicability of our
model and a comparison to the previous theory \cite{BWH}.

\section{Simple case: DNA is short}

\subsection{Model and approach}

We assume that within some volume $v$ a number of short, rigid (double
helix) DNA molecules of length $l$ are confined, among which only
one piece of DNA contains a target site of size $b$. We call
this molecule the specific DNA while others are called nonspecific
DNA. The system considered here is equivalent to an \emph{in vitro}
experiment with specific DNA concentration $1/v$ and much larger
nonspecific DNA concentration $N$.

We further assume that a protein can be adsorbed non-specifically on
DNA, and that the nonspecific adsorption energy $\epsilon$, or the
corresponding constant $y = \exp(\epsilon/k_BT)$, is the same
everywhere on the DNA and does not depend on the DNA sequence. The
only exception is at the target site on the specific DNA, where the binding
energy is much larger. We assume that every protein has two sites
capable of binding to DNA, so that the protein can be bound
to two DNA molecules at the same time.

A non-specifically bound protein is assumed to diffuse (slide)
along DNA with the diffusion coefficient $D_1$, while protein
dissolved in the surrounding water diffuses in 3D with diffusion
coefficient $D_3$. In the simplest version of the theory, we assume
$D_1 = D_3 = D$. While the protein is diffusing, the DNA molecule itself
diffuses through water with diffusion coefficient $D_t$. Following
the Stokes-Einstein relation, $D_t \sim D(b/l)$, where $b$ is the
size of the protein.

The quantity of our interest is the mean time $t$ needed for the
target site to be found by a protein. We want to look at the
situation in terms of a single protein diffusing to its target. In
this view, one should imagine that a protein molecule is initially
introduced into a random place within volume $v$ (thus the protein
concentration $c$ is $1/v$), and then ask how fast the protein
diffuses to its target site on the specific DNA. In order to compare
the predicted time $t$ to the Smoluchowski time $t_s = 1/J_s =
1/4\pi Dcb$, we shall mainly look at the acceleration rate
\be \frac{t_s}{t} = \frac{1}{t (4\pi Dcb)} \sim \frac{v}{t Db}
\label{eq:ratio}. \ee

We note that in our scaling theory we drop away both all numerical factors and
all logarithmic correction factors, which exist in the problem because it
deals with strongly elongated cylinders. In this context, we will use the symbol $``\sim"$ to mean ``equal up to a numerical coefficient of order one", while symbols $<$ and $>$ mean $<<$ and $>>$, respectively.  Along with these
simplifications, we also make several assumptions driven by pure
desire to make formulae simpler and to clarify major physical ideas.
We assume that all ``microscopic" length scales are of the same
order, namely, the target size $b$: protein diameter, double
helical DNA diameter, and the distance from DNA at which nonspecific
absorbtion takes place are all considered to be roughly equal to $b$.

\subsection{Search time}

\begin{figure*}[htb]
\centering
\includegraphics[width=0.95\textwidth]{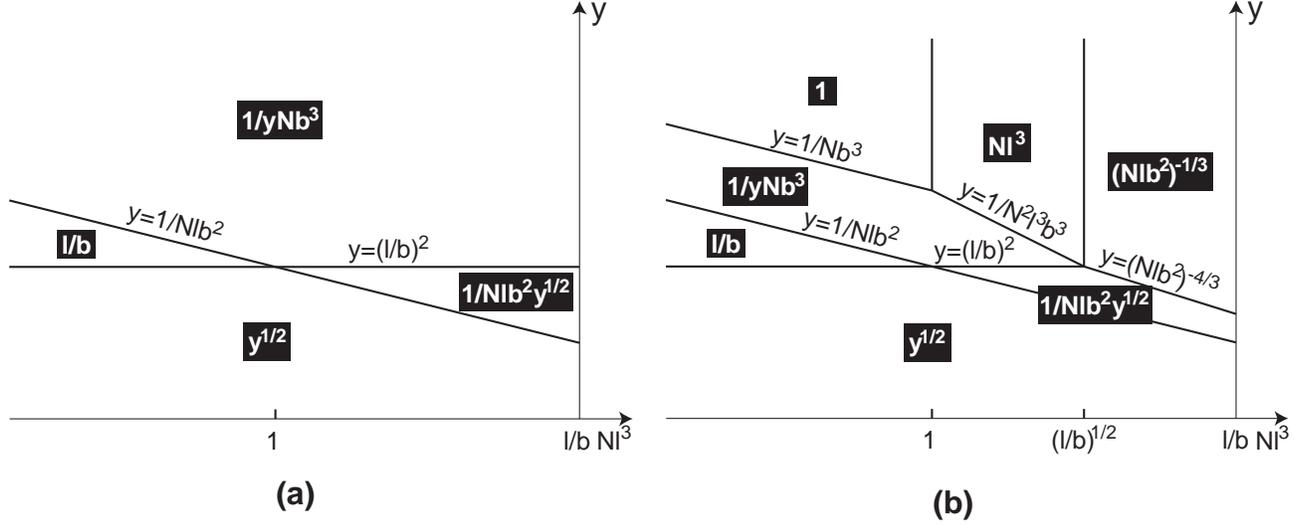}
\caption{``Phase diagram" for the acceleration rate $t_s/t$ in the
plane of $y$ and $Nl^3$, where $l$ is held constant. Both the $y$
and $Nl^3$ axes are in the logarithmic scale. The ratio $t_s/t$ is
shown in black on the background of each region. (a) shows scaling
dependencies without inter-DNA transfer; (b) gives results with
inter-DNA transfer, where $t_s/t$ saturates at large
$y$.}\label{fig:searchrate}
\end{figure*}
Let us imagine for a moment that there is no intersegment transfer,
as is the case for a protein with only one nonspecific DNA binding
site. One protein is introduced into the volume $v$. The ensuing search
process for the given single protein consists of tours of 1D sliding
along the nonspecific DNA followed by 3D diffusion in water,
followed by 1D sliding, and so on. On its way to the target on the
specific DNA, the protein will go through many adsorption and
desorption cycles, and therefore the ratio of the typical time for the
protein to be adsorbed, $t_a$, and desorbed, $t_d$, in a cycle should simply
follow the equilibrium Boltzmann statistics:
\be \frac{t_a}{t_d} \sim y(Nlb^2). \label{eq:time}\ee
The diffusion time in water per cycle $t_d$ can be estimated as the
time a protein needs to find a DNA molecule and bind nonspecifically
to it. Using Eq. (\ref{eq:smoluchowski}), $t_d \sim 1/(1/v)D(Nv)l
\sim 1/NDl$, where $1/v$ stands for the protein concentration $c$ and
$Nv$ is the number of DNA molecules in the volume $v$. As a
result $t_a \sim t_dy(Nlb^2) \sim y(b^2/D)$.

\begin{figure}[htb]
\centering
\includegraphics[width=0.45 \textwidth]{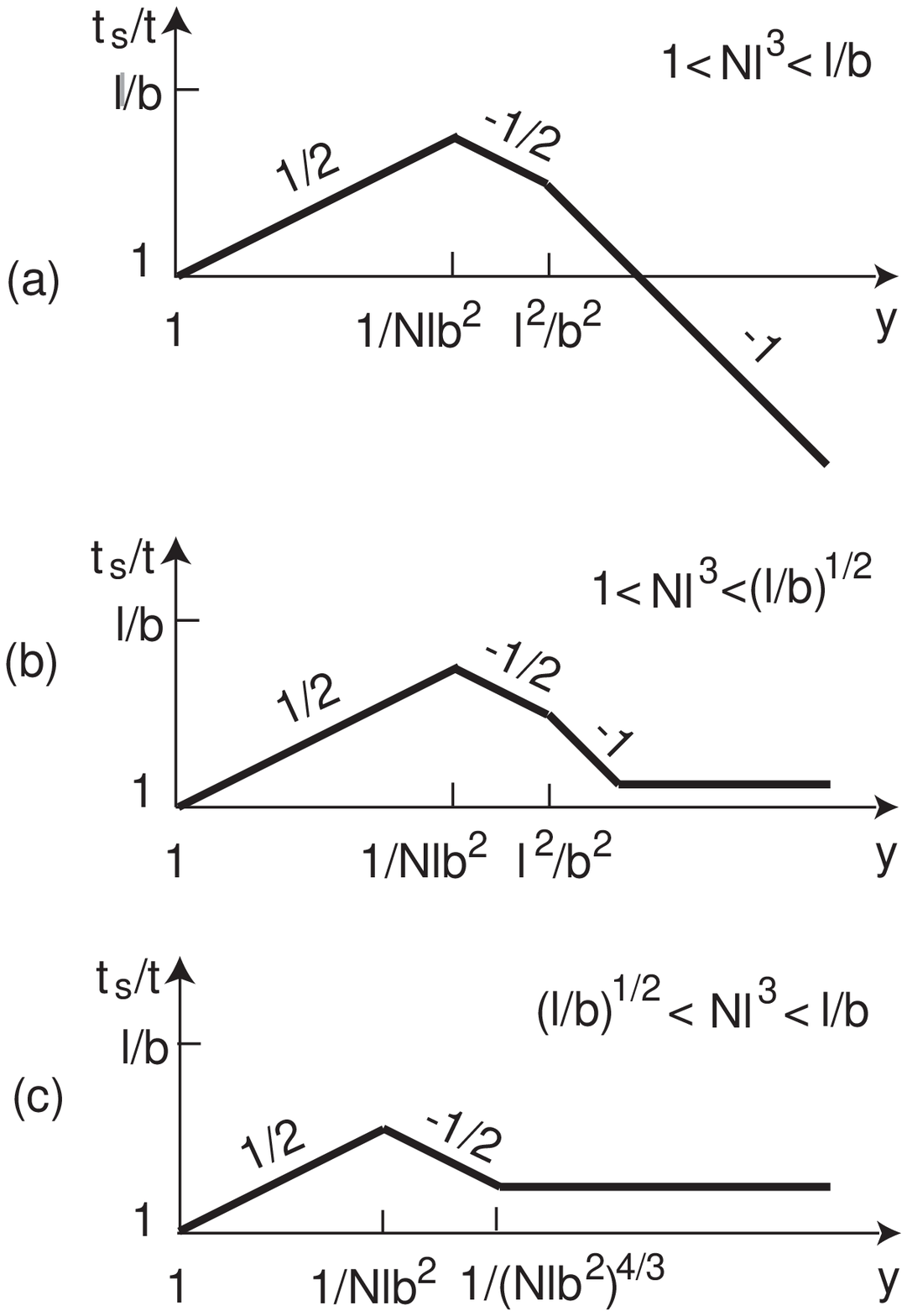}
\caption{Schematic dependencies of acceleration rate on $y$ for a
semi-dilute ($Nl^3>1$) solution of short DNA pieces. (a) without
inter DNA transfer; (b) and (c) with inter DNA transfer. The
fraction next to each curve shows its slope (the power dependence of
$t_s/t$ on $y$).} \label{fig:searchrateyy}
\end{figure}

Let $x$ be the average length of DNA searched by the protein per cycle. Then
in order to find the specific site (target) among the total
$Nvl/b$ sites on DNA, the protein should perform such searching
cycles roughly $Nvl/x$ times. Therefore the search time is given by
\be t \sim \frac{Nvl}{x}(t_a + t_d) = \frac{v}{Dx}(1+yNlb^2). \ee
Plugging $t$ into Eq. (\ref{eq:ratio}), we obtain the acceleration
rate
\be \frac{t_s}{t} \sim \frac{1}{1+yNlb^2}\frac{x}{b}.
\label{eq:ratio1} \ee
We can consider two limiting cases to find expressions for $x$. At
$y<l^2/b^2$, $x \sim (Dt_a)^{1/2} \sim y^{1/2}b$ is just the sliding
distance of the protein on one DNA molecule, while at $y > l^2/b^2$,
$x$ is limited to the total length of DNA $l$. There are also two
limiting cases for the denominator of Eq. (\ref{eq:ratio1}). When
$y$ is relatively small so that $y < 1/Nlb^2$, i. e. the protein
spends most of its time desorbed in water, the first term dominates.
At $y > 1/Nlb^2$, the protein spends most of its time adsorbed and
the second term dominates. As a result we obtain four scaling
regimes shown in the phase diagram of Fig. \ref{fig:searchrate}(a).
We terminate the phase diagram at the concentration $Nl^3=l/b$
because in a denser system liquid crystalline nematic ordering of
DNA molecules becomes likely. The dependencies of the acceleration
rate on $y$ for a semi-dilute solution of short DNA pieces are
plotted in Fig. \ref{fig:searchrateyy}(a). The search rate is shown
to increase at first due to the increase of $x$, and then decrease
due to the fact that at large $y$ the protein spends most of the
time adsorbed on nonspecific DNA molecules, which slows down the
diffusion to the target.

Now let us move on to the case of a protein with two DNA binding
sites. When a piece of DNA with an adsorbed protein collides
with another DNA molecule, the protein has some probability to move
directly to the new molecule. If the inter-DNA transfer is faster than
the dissociation of protein into water, i.e. if the average time $\tau_t$
required for a protein to be transferred from one piece of DNA to another
is shorter than the adsorption time $t_a$, then the
protein can explore several DNA molecules during $t_a$. As a result,
the protein can visit a large number of different sites during adsorption and the
efficiency of 1D search on DNA is greatly enhanced. We find below
that at large $y$ when $\tau_t < t_a$ and inter-DNA transfer
dominates, the protein already spends most of the time adsorbed and
$t_a
> t_d$. Therefore we neglect the time spent in water $t_d$ and
redefine $x/b$ as the number of different sites explored on the same
DNA during $\tau_t$. The search time can then be estimated as
\be t \sim \frac{Nvl}{x}\tau_t, \label{eq:stime}\ee
so that we obtain the acceleration rate
\be \frac{t_s}{t} \sim \frac{x}{NDlb\tau_t}.\ee
The results are shown in the diagram Fig. \ref{fig:searchrate}(b).
At large $y$ the acceleration rate stops decreasing with $y$ and
saturates.

We begin explaining our results by calculating the most important
quantity of our theory: $\tau_t$. For a dilute solution of DNA
molecules with $Nl^3 < 1$, one can use Eq. (\ref{eq:smoluchowski}) to
find the time for a given DNA molecule to enter the spherical region
occupied by another piece of DNA by replacing $D_3$ and $b$ by the
DNA diffusion coefficient $D(b/l)$ and the length $l$. The result is
$1/D(b/l)Nl = 1/DNb$. When the given DNA molecule enters the sphere of another molecule and
diffuses over distance $l$, on average every site on the DNA has a chance to
collide with the second DNA before it leaves the sphere.
As a result, a protein adsorbed on one DNA can essentially always get
transferred to the new one during a collision. Since in a dilute solution the diffusion
time to find such a sphere containing a second DNA piece, $1/DNb$, is larger than the diffusion time within the sphere, $l^2/D(b/l) \sim l^3/Db$, the transfer waiting
time $\tau_t$ is the order of $1/DNb$. Because $D\tau_t > l^2$, the
protein searches $l/b$ different sites during $\tau_t$ and $x \sim
l$. Using Eq. (\ref{eq:stime}), we obtain the search time
\be t\sim \frac{Nvl}{l}\tau_t \sim \frac{v}{Db} \sim t_s.
\label{eq:timeshort1} \ee

\begin{figure}[htb]
\centering
\includegraphics[width=0.25 \textwidth]{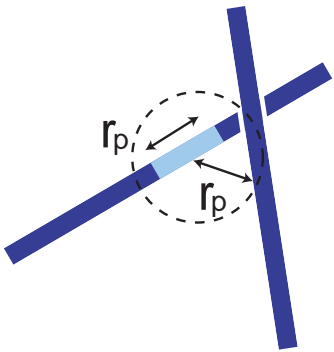}
\caption{Collision of a DNA molecule with its nearest neighbor
at distance $r_p$ (other DNA molecules are not shown). }
\label{fig:neighbor}
\end{figure}

When $Nl^3 >1$, the spheres containing individual DNA molecules
strongly overlap. In such a semi-dilute solution, the first collision
for a given DNA molecule happens when it diffuses over the nearest
neighbor distance $r_p$. One can find $r_p$ by constructing an imaginary
cylinder with radius $r_p$ around each DNA molecule, where the length of the molecule serves as the cylinder's axis.
Because the excluded volume of a cylinder is $\sim l^2r_p$, the radius
$r_p$ should satisfy $Nl^2r_p \sim 1$ and thus scale as $1/Nl^2$.
During time
\be \tau \sim \frac{r_p^2}{D(b/l)} \sim \frac{1}{DN^2l^3b}, \ee
the DNA diffuses over a distance $r_p$, giving every site on some segment
of the DNA of length $r_p$ the opportunity to collide once with the nearest-neighboring
DNA (see Fig. \ref{fig:neighbor}). After time  $\tau$,
the diffusing DNA and its neighbors have moved around enough that the nearest-neighboring region (shown by a dashed circle) may be considered to have shifted to a random place on the DNA.

Let us assume that the protein has just arrived at some place on the
given DNA molecule. In order to be transferred to another DNA within time $\tau$, the protein must
reach the segment of length $r_p$ (see Fig. \ref{fig:neighbor})
during $\tau$. Since the typical distance between the adsorbed
protein and the nearest neighboring region is just proportional to the DNA length
$l$, the protein will change molecules during $\tau$ when $D\tau >l^2$,
or $Nl^3 < (l/b)^{1/2}$. Therefore $\tau_t \sim \tau$ and we
obtain the search time
\be t \sim Nv\tau_t \sim \frac{1}{Nl^3}\frac{v}{Db} \sim
\frac{t_s}{Nl^3}, \label{eq:timeshort2}\ee
from which we can see that the search rate saturates $Nl^3$ times
faster than the Smoluchowski rate, and that the acceleration rate
grows with DNA concentration since denser solution makes inter-DNA
transfer easier.

When $Nl^3 > (l/b)^{1/2}$, the 1D sliding distance of protein on a
single DNA molecule during $\tau$ is $x \sim (D\tau)^{1/2} < l$. Therefore, the
probability for the protein to reach the nearest neighboring region
on the DNA during $\tau$ is $x/l \sim (l/b)^{1/2}/Nl^3 <1$. In this
case the transfer waiting time $\tau_t > \tau$, and it should be
calculated self-consistently. During $\tau_t$ the sliding distance $x$
of the protein is $(D\tau_t)^{1/2}$, so the probability for the
protein to reach a specified nearest neighboring region is on the order
of $(D\tau_t)^{1/2}/l$. Since the nearest-neighboring region changes to a random place on the DNA after $\tau$, there are $\tau_t/\tau$ such regions during time $\tau$. Therefore the probability for the protein to reach any one of these regions and then get transferred should
satisfy
\be \frac{(D\tau_t)^{1/2}}{l}\frac{\tau_t}{\tau} \sim 1. \ee
As a result,
\be \tau_t \sim \frac{1}{D(N^2l^2b)^{2/3}}, \label{eq:tautshort3}\ee
and the search time is given by
\be t \sim \frac{Nvl}{(D\tau_t)^{1/2}}\tau_t \sim
(Nlb^2)^{1/3}\frac{v}{Db} \sim (Nlb^2)^{1/3}t_s.
\label{eq:timeshort3}\ee

The equations for the crossover lines at large $y$, shown in Fig.
\ref{fig:searchrate}(b), are obtained by equating $\tau_t$ to $t_a$.
This condition determines the range of parameters for which
intersegment transfer takes over, i.e. when the time it takes the
protein to transfer between DNA molecules is much shorter than the
time the protein spends adsorbed on the DNA then we can say that
intersegment transfer is the dominant mechanism. The dependencies of
the acceleration rate on $y$ for semi-dilute DNA concentrations with
$Nl^3 > 1$ are schematically plotted in Fig. \ref{fig:searchrateyy}
(b) and (c). For the purpose of comparison, we also show the
dependencies for proteins with a single binding site in Fig.
\ref{fig:searchrateyy} (a).

A new feature shown by Fig. \ref{fig:searchrateyy} (b) and (c) is
that, for proteins with two DNA binding sites, inter-DNA transfer
stops the search rate from decreasing and causes it to saturate at
large $y$. It can be shown from equations (\ref{eq:timeshort1}),
(\ref{eq:timeshort2}) and (\ref{eq:timeshort3}) that the
acceleration rate is constant and $\sim 1$ when the solution is
dilute and $y$ is large. The acceleration rate begins to grow as the
concentration is increased past $Nl^3 \sim 1$, peaking when  $Nl^3
\sim (l/b)^{1/2}$ and achieving a maximum value of $(l/b)^{1/2}$.
After the peak, it decreases again and reaches $(l/b)^{1/3}$ when
$Nl^3 \sim l/b$.

Before we move on to next section, we should emphasize that in our
calculation we have completely neglected the energy barrier associated with breaking the
double-bound state. We have assumed the barrier to be so small that the lifetime of the double-bound state is a small correction to the above
calculated $\tau_t$. The search time we have found is therefore the lower limit
which can be achieved with the help of intersegment transfer. In
Sec. IV, we will return to this issue in more detail.

\subsection{Dissociation rate}

Since in experiments \cite{Fried,Lieberman,Iwahara} the role of
intersegment is inferred from measuring the dissociation rate of the
prepared protein-DNA complex, in this section we calculate this rate
for a protein adsorbed on a nonspecific piece of DNA dissociating to
other nonspecific DNA pieces via inter-DNA transfer. \footnote{Our
model serves as a simple generalization of experimental systems
where the protein is specifically adsorbed to its target on the DNA
in the complex and the DNA free in solution can be specific or
nonspecific \cite{Fried,Lieberman,Iwahara}. We argue that this
generalization does not change the main feature of the problem,
which is determined by the frequency at which the free DNA molecules
collide with the protein-DNA complex. The difference lies in the
transfer probability per collision. On the nonspecific DNA, the
protein can slide freely. In contrast, protein on the specific DNA
spends most of the time adsorbed to its target. Thus, to experience
a transfer the complex should collide with another piece of DNA
exactly at the position of target, which results in smaller transfer
probability. However the specifically adsorbed protein can first
slide into nonspecific sites and then dissociate into the bulk
solution or transfer to other DNA molecules \cite{Shimamoto}. In
this way the decreased transfer probability is somewhat compensated,
and it becomes closer to the case of dissociation from nonspecific
DNA. More importantly, the process of target search involves the
protein making direct transfers between nonspecific segments of DNA,
so we prefer to study the dissociation rate for this case.}

\begin{figure}[htb]
\centering
\includegraphics[width=0.40 \textwidth]{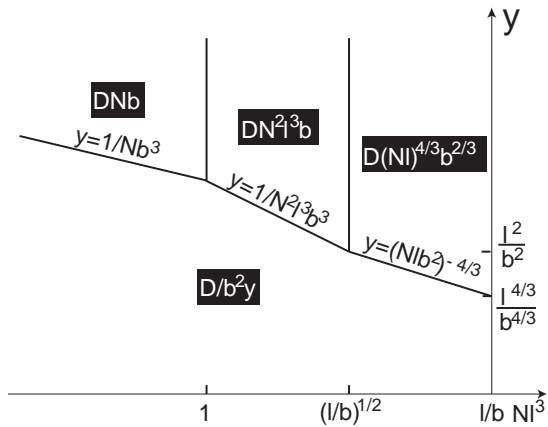}
\caption{``Phase diagram" for the dissociation rate under the
influence of inter DNA transfer.} \label{fig:drate}
\end{figure}

The calculation is quite straightforward and the results are
presented in the phase diagram of Fig. \ref{fig:drate}. The apparent
dissociation rate is just $1/t_a +1/\tau_t$, where each term
represents a possible relaxation process undergone by the adsorbed
protein: either dissociation to water or intersegment transfer to
another piece of DNA. The faster process dominates the rate. Since
the dissociation rate to water decreases with the adsorption
strength $y$ and the intersegment transfer rate grows with the
nonspecific DNA concentration, intersegment transfer dominates the
apparent dissociation rate at relatively large $y$ and $N$. We find
that the enhanced dissociation rate grows linearly with nonspecific
DNA concentration $N$ when the solution is dilute, in agreement with
the experiments \cite{Fried,Lieberman,Iwahara}. In semi-dilute
solution, however, the dissociation rate has power law dependence on
$N$, with power equal to either $2$ or $4/3$.

\section{DNA is a globule}

After exploring the particular role of DNA motion and protein
intersegment transfer for short DNA pieces, we are well prepared to
generalize the above results to the more realistic case of globular
DNA (see Fig. \ref{fig:globule}). We focus here on cases with large
$y$, where the mechanism of intersegment transfer is important. The
results for acceleration rate at small $y$, where intersegment
transfer does not help much, can be found in Ref. \cite{BPJ}.

We assume that within some volume $v$, a double helical DNA with contour
length $L$ and persistence length $p \gg b$ is confined. We
disregard the excluded volume of DNA, considering the DNA coil to be
Gaussian and not a swollen coil, described by the Flory index
$3/5$. This is a reasonable approximation for most realistic cases.
Indeed, for many real DNA molecules such as $\lambda$-DNA,
it is justified because of the large persistence length-to-diameter
ratio of the double helix: excluded volume in the coil remains
unimportant up to DNA length about $L < p^3/b^2$ (as much as
$100000$ base pairs under normal ionic conditions). When the DNA is
very long for a given volume, specifically, when the gaussian coil
size $(Lp)^{1/2}
> v^{1/3}$, it cannot remain a Gaussian coil, but must fold back to
make several smaller overlapping coils. In other words, it must be a
globule which locally resembles a transient network.

In order to simplify our calculation, we can approximate the DNA as
a series of freely-jointed straight segments (rods), each with persistence length
$p$.  We further restrict our study to a globule so dense that the
spheres containing each rod strongly overlap (Fig.
\ref{fig:globule}). Except for the connectivity, the globule is
quite similar to a semi-dilute solution of short straight DNA pieces
of length $p$ and concentration $N = (L/p)/v$ satisfying $Np^3 > 1$.
In this case the diffusion distance $r_p$ for a given rod to
experience its first collision with another rod, which may be close in space but
far removed along the DNA contour, is shorter than its length $p$. As a
result, one can disregard the correlation of motion between connected
rods and treat the motion of each rod over the short distance
$r_p$ separately as a normal diffusion process with diffusion coefficient
$D(b/p)$.

Let us first look at a simple case where the 1D sliding distance $x$
for a protein on a single DNA rod is shorter than the chain length
$p$. As before, we consider $x$ to be the distance traveled by the protein
within a time $\tau_t$, the average waiting time before a protein is transferred
from one DNA rod to another, uncorrelated rod. In this situation, the protein does not feel the connection between rods. Therefore, we can simply use the result for short
DNA pieces, replacing the length $l$ by $p$ and using the rod
concentration $N = (L/p)/v$. Then Eq. (\ref{eq:timeshort3}) gives
the search time
\be t \sim (Npb^2)^{1/3}(v/Db) \sim (Lb^2/v)^{1/3}t_s.
\label{eq:timelong1}\ee
From Eq. (\ref{eq:tautshort3}), we find $\tau_t \sim
1/D(N^2p^2b)^{2/3}$ and thus $x \sim (D\tau_t)^{1/2} \sim
1/(N^2p^2b)^{1/3}$. So the condition $x < p$ is fulfilled when $Np^3
> (p/b)^{1/2}$ or $L>(v/b^2)(b/p)^{3/2}$. Furthermore,
to avoid the liquid crystalline nematic ordering of DNA chains, we
assume that $Np^3<p/b$ or $L<v/pb$.

When the concentration of DNA rods is small enough that it falls
within the range $1 < Np^3 < (p/b)^{1/2}$, the separation between
DNA rods becomes large. Therefore, the time between collisions
increases. As a result, the transfer waiting time $\tau_t$ grows and
the 1D sliding distance of the protein $x$ becomes larger than $p$.
In this case, one should be careful in calculating the DNA diffusion
distance that results in the first collision between DNA rods. It is
no longer equal to the nearest neighbor distance $r_p \sim 1/Np^2$
between DNA rods of length $p$. To find this distance, let us
concentrate on the continuous piece of length $x > p$, which spans
several rods. The shortest distance from this piece of DNA to
another similar piece is realized at only one of its constituent
rods. The first collision that could result in transfer of the
protein happens only when this particular rod diffuses over the
$x$-dependent nearest neighbor distance $r(x) \sim r_pp/x \sim
1/Npx$. During time $\tau(x) \sim r^2(x)/D(b/p)$, on average each
DNA piece of length $x$ experiences a collision, and the protein
slides a distance $x$ across the DNA. Thus, the waiting time for a
protein to be transferred to another, uncorrelated DNA piece $\tau_t
\sim \tau(x) \sim 1/DN^2x^2pb$ should be equal to the 1D sliding
time $x^2/D$ of the protein on a single piece. This self-consistent
calculation gives $x \sim (1/Np)^{1/2}(p/b)^{1/4}$ and $\tau_t \sim
(1/DNp)(p/b)^{1/2}$. We therefore obtain the search time
\be t \sim \frac{L}{x}\tau_t \sim
(Np^3)^{1/2}\left(\frac{b}{p}\right)^{3/4}\frac{v}{Db} \sim
\left(\frac{Lb^2}{v}\right)^{1/2}\left(\frac{p}{b}\right)^{1/4}t_s.
\label{eq:timelong2}\ee

As explained in Ref. \cite{BPJ}, without intersegment transfer,
large values of $y$ result in the protein spending most of its time
adsorbed on DNA far from the target site. The result is that the
search time saturates at $L^2/D \sim (L^2b/v)t_s$, which is a huge
deceleration compared to the Smoluchowski time. From Eqs.
(\ref{eq:timelong1}) and (\ref{eq:timelong2}), one can easily find
that at large $y$ the search time is greatly reduced below $t_s$ by
the combination of 1D sliding, intersegment transfer and DNA motion.
Correspondingly, the acceleration rate is enhanced and can be larger
than $1$, as shown by the solid line in Fig. \ref{fig:globulerate}.
These results remain qualitatively correct for a sparser globule
with $p/v^{1/3}<Np^3<1$ or $v^{2/3}/p<L<v/p^2$, where the typical
mesh size of the transient network is longer than $p$ and thus the
piece of DNA inside each mesh is not straight as shown in Fig.
\ref{fig:globule} but rather a small Gaussian coil. To fully account
for this kind of geometry, however, one should consider a more
complicated correlated segmental diffusion of DNA, and this is
beyond the scope of the current paper.

\begin{figure}[htb]
\centering
\includegraphics[width=0.40 \textwidth]{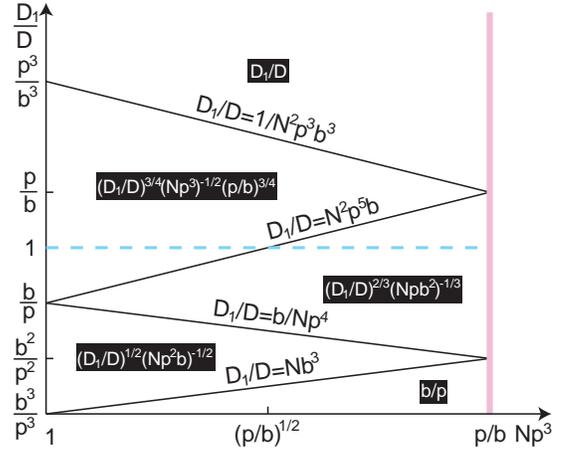}
\caption{``Phase diagram" for the acceleration rate at large $y$ on
globular DNA, where the intersegment transfer plays an important
role in the target search.} \label{fig:done}
\end{figure}

Until now, we assumed that $D_1 = D_3 = D$, where $D_1$ and $D_3$
are the diffusion coefficients of protein on DNA and in water,
respectively. In fact, the random sequence of DNA and the resulting
sequence-dependent nonspecific adsorption energy most likely
produces $D_1 < D_3$. To illustrate the role of 1D sliding in
conjunction with intersegment transfer, we fix $D_3=D$ and calculate
the acceleration rate for various values of $D_1$ following the
methods explained above. The results for large $y$, where the
intersegment transfer plays an important role, are shown in the
plane of $D_1/D$ and $Np^3$ in Fig. \ref{fig:done}. The dashed line
corresponds to $D_1=D_3=D$. We find that the acceleration rate grows
as $(D_1/D)^S$ with the index $S$ increasing from $1/2$ to $1$.

\section{Discussion}

In our theory, we completely neglect the effect of the energy
barrier $\epsilon^*$ associated with breaking the double-bound state
and reverting to a single-bound state. Our results are therefore an
upper estimate of the effect of intersegment transfer. On the other
hand, a na\"{\i}ve guess of the barrier height is $\epsilon^* =
\epsilon$, since to break one of the two contacts the protein has to
pay the adsorption energy per binding site on one side. If this were
true, the protein would be trapped in the double-bound state for the
adsorption time $t_a$, and therefore the inter-DNA transfer could
not do better job in accelerating the dissociation of the protein
from the protein-DNA complex than desorption into water. As a
result, adding DNA into the solution of a protein-DNA complex would
not increase the dissociation rate of the protein, which clearly
contradicts the \emph{in vitro} experiments on various proteins and
DNA molecules \cite{Fried,Lieberman,Iwahara}. This suggests that in
the double-bound state, the binding strength per binding site
$\epsilon^* < \epsilon$, which could be a result of the excluded
volume of close DNA molecules or the Coulomb repulsion between them.

The experiment \cite{Fried} showed that the dissociation rate
increases linearly with the nonspecific DNA concentration and
saturates at large concentrations. This implies that at small DNA
concentration, the dissociation rate is limited by the diffusion of
nonspecific DNA molecules and the resulting collisions that induce
inter-DNA transfer. As the DNA concentration is increased, the
energy barrier for releasing the protein from the double-bound state
becomes the bottleneck of the dissociation. Since the lifetime of
the double-bound state does not depend on the nonspecific DNA
concentration, the dissociation rate saturates. Having $\epsilon^* <
\epsilon$ in mind, one can show that our theory is valid if the
transfer waiting time $\tau_t$ is larger than the lifetime of the
double-bound state. We can estimate this lifetime as the product of
the characteristic time scale $b^2/D$ and the binding strength per
site in the double-bound state $y^* \sim \exp(\epsilon^*/k_BT)$.
Thus our theory works when $y^* (b^2/D) < \tau_t$. When $y^*(b^2/D)
> \tau_t$, our main idea is still correct, however one should
replace $\tau_t$ by the lifetime of the double-bound state
$(b^2/D)y^*$ and repeat a similar analysis. The acceleration rate
will be diminished as a result but will remain much larger than in
the case without intersegment transfer.

The above discussion of $\epsilon^*$ assumes that the double-bound
state does not affect the equilibrium Boltzmann statistics
represented by Eq. (\ref{eq:time}). This places an additional
restriction on $\epsilon^*$. The energy of the double-bound state is
$2\epsilon^*$. If one were to take a snapshot of the solution of
short DNA pieces at a given time, the number of DNA contacts (where
two DNA collide) per DNA strand is on the order of $Nl^2b$, where
$l^2b$ represents the excluded volume of a rod-like DNA. Then the
limitation on $y^*$ can be expressed as
$\exp(2\epsilon^*/k_BT)(Nl^2b)(b^3) < ylb^2$ or $(y^*)^2 < y/Nlb^2$.

Let us now compare our work with the treatment of intersegment
transfer in Ref. \cite{BWH}. While our work combines both mechanisms
of 1D sliding and intersegment transfer, the Ref. \cite{BWH} treats
them separately. Neglecting the mechanism of protein sliding in a
description of intersegment transfer results in a huge
overestimation of the collision time $\tau$ and the subsequent
transfer time $\tau_t$. Indeed one can see from Fig. \ref{fig:done}
that if the protein cannot move on DNA, the acceleration rate is
$b/p$, which is much smaller than the acceleration rate at
$D_1=D_3=D$. Equivalently, neglecting intersegment transfer results
in overestimation of the sliding time, sliding distance and search
time. In the later review of Ref. \cite{Berg3}, the interplay
between sliding and intersegment transfer was taken into account.
Qualitatively, the conclusions agree with our results, however the
dependence of the intersegment transfer rate on the characteristics
of DNA geometry, DNA motion, concentration and the nonspecific
adsorption strength of protein to DNA $y$ was not calculated.

Finally, we note that our theory can be easily adapted to study the
effective diffusion rate of a protein through a solution of polymers
like DNA. This problem was studied in Ref. \cite{Berg2}, assuming
$D_1=0$. Following the ideas of our paper, one can expand on this study to account for the
``constructive interference" of 1D sliding and intersegment transfer
of protein, which was not addressed in Ref. \cite{Berg2}. As with
target search, intersegment transfer enhances the macroscopic
diffusion coefficient of proteins at large $y$, where the protein
spends most of its time adsorbed on DNA. We can consider a solution of short DNA
molecules, where without intersegment transfer the effective diffusion
coefficient of the protein is decreased by nonspecific adsorption
to DNA and eventually saturates at the DNA diffusion coefficient
$D(b/l)$. In a dilute solution, intersegment transfer does not assist the macroscopic
diffusion of proteins, since
each DNA molecule is far removed from other molecules and therefore the macroscopic
displacement of protein is determined mainly by the motion of the DNA.
In a semidilute solution, however, where $1 < Nl^3 < l/b$, 1D sliding on
DNA becomes important. When $1<Nl^3<(l/b)^{1/2}$, $D\tau_t>l^2$ and
the 1D sliding distance of protein during time $\tau_t$ is limited
to the length of DNA $l$. In this case $\tau_t \sim \tau$. Using Eq.
(\ref{eq:stime}), the effective diffusion coefficient is obtained as
$l^2/\tau_t \sim N^2l^6D(b/l)$. At higher densities when $(l/b)^{1/2} < Nl^3 < l/b$,
$D\tau_t < l^2$. As a result, the nonspecific adsorption of protein
on DNA does not hinder the diffusion of protein at all and the
macroscopic diffusion coefficient is just $D$. For $D_1 \neq D_3$,
a similar analysis can be performed.

One further application of our theory is to the problem of dynamic (stirred)
percolation, e.g., the conductivity of well-conducting wires in some
insulating liquid. It is well known that if the wires are randomly
frozen in the liquid, the conductivity vanishes below the
percolation threshold \cite{Boris}. However, because of the
diffusion of wires in the liquid, the charge carriers are not
trapped within finite clusters of wires. Instead, they can hop from one
wire to another when the wires approach close to each other. This results in a
finite conductivity below the percolation threshold
\cite{percolation}. For such systems, one can find the macroscopic
diffusion coefficient of the charge carriers and then map it to the
effective conductivity of the system.

\acknowledgements We are grateful to A. Yu. Grosberg and M. G. Fried
for helpful discussions. We acknowledge B. Skinner's kind help in
preparing the manuscript. TH acknowledges support of the Anatoly
Larkin Fellowship.


\end{document}